\documentclass[pra,twocolumn,nofootinbib,eqsecnum,floatfix]{revtex4}
\usepackage{bm} 
\usepackage{amsmath}

\begin{document}

\title{{\bf Typicality Defended}
\thanks{Alberta-Thy-10-07, arXiv:yymm.nnnn}}

\author{Don N. Page}
\email{don@phys.ualberta.ca}

\affiliation{Institute for Theoretical Physics\\
Department of Physics, University of Alberta\\
Room 238 CEB, 11322 -- 89 Avenue\\
Edmonton, Alberta, Canada T6G 2G7\footnote{Permanent address},}
\affiliation{Asia Pacific Center for Theoretical Physics, Pohang
790-784, Korea}

\date{2007 July 27}

\begin{abstract}

Hartle and Srednicki have argued that there is no observational evidence
favoring our typicality.  Here it is shown that such evidence does arise
from including the  {\it normalization principle\/} requirement that the
sum of the likelihoods for all possible observations is normalized to
unity in each theory.

\end{abstract}


\maketitle

\section{Introduction}

As Hartle and Srednicki \cite{HS} correctly note, ``An increasingly
common kind of reasoning in fundamental cosmology starts from an
assumption that some property of human observers is typical in some class
$\cal C$ of objects in the universe,'' for example in
\cite{BLL94,Vil95,DKS02,Page06a,BF06,Page06b,Linde06,Page06c,Vil06,
Page06d,Banks07} which they cite later.  They go on to claim \cite{HS}
that ``it is perfectly possible (and not necessarily unlikely) for us to
live in a universe in which we are not typical.''  While I agree that it
is perfectly possible, in this paper I shall argue that it would be
unlikely when one properly normalizes the likelihoods.

That is, I shall argue that within each possible theory of the universe,
the likelihood would be small that we are atypical, though one could
assign a sufficiently high prior probability to a theory in which we are
atypical to overcome this small likelihood.  That is, the theory might
have such high {\it a priori\/} probability that after a Bayesian analysis
it could have the highest {\it a posteriori\/} probability even though it
makes us atypical and unlikely.  However, purely from the likelihoods,
properly normalized, typicality is favored, contrary to what Hartle and
Srednicki conclude when they do not require that the sum of the
likelihoods for all possible observations sum to unity.

A key issue to be discussed below is how likelihoods are to be defined by
a theory, which seems to lie at the heart of Hartle and Srednicki's
disagreement with calculations favoring typicality.  Another key issue is
the gap between the first sentence of their conclusion (v), ``We have
data that we exist in the universe, but we have no evidence that we have
been selected by some random process,'' with which I agree, and the
second sentence of that conclusion, ``We should not calculate as though
we were,'' with which I disagree.

Before getting into these points of disagreement, it may be helpful to
list points of agreement.

I do agree with Hartle and Srednicki's conclusion (i), ``A theory is not
incorrect merely because it predicts that we are atypical.''  Low
typicality merely implies low likelihood, but one must also consider the
prior probability assigned to the theory.

I also agree with conclusion (iii), ``No part of our data should be
neglected in the process of discriminating between competing theories
unless it can be demonstrated that the relevant probabilities are
insensitive to it.''

Strictly speaking, I also agree with conclusion (vi), whose first
sentence is, ``In a fundamental theory of quantum cosmology, there is no
need for  any assumption of typicality to predict what we might see.''  I
would say that all we need to do is consider theories that predict
correctly normalized likelihoods for all possible observations, and then
the typicality will be automatically reflected in these likelihoods.

I furthermore strongly agree with Hartle and Srednicki in using Bayesian
probability theory \cite{Sred05,Jaynes,App04}, with its prior probability
$P(T_i)$ for each theory $T_i$, its likelihoods $P(D_j|T_i)$ or
conditional probabilities for each possible data set $D_j$ given the
theory $T_i$, and its posterior probabilities $P(T_i|D_j)$ or reverse
conditional probabilities for the theories $T_i$ given the particular
data set $D_j$ that is obtained.  These posterior probabilities are given
in terms of the prior probabilities and the likelihoods by Bayes'
theorem,
\begin{equation}
P(T_i|D_j) = \frac{P(D_j|T_i) P(T_i)}{\sum_i  P(D_j|T_i) P(T_i)}  . 
\label{bayesthm}
\end{equation}

In particular, I concur with Hartle and Srednicki that Bayesian analysis
provides a framework for distinguishing ``facts, logical deduction, and
prejudices.''  As they nicely express it, ``Data are the domain of facts,
likelihoods are the domain of logical deduction, and the priors are the
domain of theoretical prejudice.'' 

In this Bayesian approach, my key difference from Hartle and Srednicki is
that I propose that one follow the {\it normalization principle\/}:  One
should only consider theories that each give likelihoods summing to unity
for all possible data sets,
\begin{equation}
\sum_j P(D_j|T_i) = 1  . 
\label{norm}
\end{equation}

I shall take an atypical observation to be an observed data set that has
an anomalously low likelihood, so that atypical observations are
unlikely, giving small weights in Bayes' theorem.  (I think of
observations as being more fundamental than observers and hence shall
focus on typical or atypical observations rather than on typical or
atypical observers, but one could define an atypical observer as one who
makes atypical observations.)

Here I am not using the technical definition of typicality I have
proposed in \cite{SQM1,SQM2,Page-in-Carr}, which after email discussions
with Srednicki \cite{Sredpriv} I realize has some problems that I shall
discuss elsewhere \cite{Pagetyp}, but the looser idea that atypical
observations are those that would be unlikely to be chosen in a random
selection from all observations predicted by the theory.  Although a
precise definition is not needed here, it might help to have the
following definition in mind as an example:

Consider all the observed data sets $D_j$ predicted by some theory $T_i$,
and rank them in decreasing order of their normalized likelihoods
$P(D_j|T_i)$.  Define the median observation $D_m$ as the one with the
smallest value of $m$ in this ordered sequence such that $\sum_{j\leq m}
P(D_j|T_i) > 1/2$.  Then one might define the typicality of any observed
data set $D_j$ in this theory as being $t_j = P(D_j|T_i)/P(D_m|T_i)$. 
For $j\leq m$, the typicality is large, $t_j \geq 1$, so that at least
half of the likelihood occurs for observed data sets with high
typicality.  Atypical observations would correspond to low typicality,
$t_j \ll 1$, and would occur only for $j > m$ and a small fraction of the
total amount of likelihood.  That is, it is unlikely that an observation
would be atypical if it were selected randomly with a probability given
by its normalized likelihood.

\section{Data}

There are different ideas for what should constitute a data set $D_j$ to
be used in a Bayesian probability analysis.  Since each theory $T_i$ is
supposed to assign a likelihood $P(D_j|T_i)$ to each data set $D_j$, it
should be the theory that defines the possible data sets.  Thus different
ideas of what the data sets are may be considered as differences in the
theories.  However, to compare different theories by a Bayesian analysis,
they should all have data sets that are members of some single
encompassing set of data sets, say $S$.  Then for any theory whose data
sets form only a proper subset of $S$, one can simply say that that
theory predicts zero likelihood for all other data sets of $S$.  In this
way we can say that each theory $T_i$ assigns a unique value to the
likelihood $P(D_j|T_i)$ for each data set $D_j$ in the full set $S$ of
such data sets.

Although my argument does not depend on which full set of data sets $S$
is chosen (so long as it is precisely defined), let me give some possible
choices.  The one that seems the most fundamental to me is the set, say
$S_1$, of all possible conscious perceptions
\cite{SQM1,SQM2,Page-in-Carr,SQM0,SQM3,SQM4,MS}.  Roughly, each individual
conscious perception is all that a conscious observer is aware of at
once, what Bostrom \cite{Bostrom} calls an observer-moment.  If this
conscious perception is regarded as a data set, the data would be the
content of that awareness.  In this $S_1$, each different possible
conscious perception would be a member, and any two perceptions with
different contents would be different data sets.

Hartle and Srednicki use an HSI, a human scientific IGUS (information
gathering and utilizing system), with the data set including ``every
scrap of information that the HSI possesses about the physical universe:
every record of every experiment, every astronomical observation of
distant galaxies, every available description of every leaf, etc., and
necessarily every piece of information about the HSI itself, its members,
and its history.''  Although they consider only the data set $D$ that our
particular HSI has (and thereby avoid the issue of normalizing the
likelihoods over all possible data sets), one can certainly consider all
such data sets, say forming the set $S_2$.  Every such data set within
$S_2$ would differ if it had different scraps of information or different
information within at least one of the scraps.

Another possible set, say $S_3$, of data sets would be the set of all
complete physical descriptions of all planets (including what is on them,
of course).  Each physical different planet would give a different data
set in this set of data sets.

Yet another possible set of data sets, say $S_4$, would be the set of all
complete descriptions of the causal past of any event of spacetime
(assuming for this that spacetime has a definite causal structure, which
is not likely to be true in quantum gravity).

For any particular Bayesian analysis, one should have a definite set $S$
of well-defined possible alternative data sets.  There should be some
parallelism between the data sets within $S$, so it would not appear
to be a good idea, for example, to use an $S$ that is the union of the
set $S_1$ of all possible conscious perceptions and the set $S_2$ of
all possible HSI data sets, since each HSI data set may contain one or
more (usually many) conscious perceptions.

Because the set $S$ of data sets must be well defined, with distinct
members (the data sets themselves), the data sets within $S$ must all be
different and can be regarded as different alternatives, different
possible observations.  By assumption, an observation (whether a
conscious perception, all the data of an HSI, or all the data of a
planet) is of a distinct data set $D_j$ with $S$, and therefore each
theory $T_i$ should assign a definite likelihood $P(D_j|T_i)$ for each
data set $D_j$.  Since the data sets are alternatives of what might be
observed, and since by assumption some particular data set actually is
observed, for each theory $T_i$ the sum of the likelihoods (the
conditional probabilities of the data sets, given the theory) should sum
to unity, the normalization condition (\ref{norm}).

\section{Likelihoods}

Each normalized likelihood $P(D_j|T_i)$ that I have discussed above may
be regarded as the probability, conditional upon the theory $T_i$, that
an observation (randomly chosen without restricting the data) gives the
data set $D_j$.  If one considers the observation to be made by an
observer (whether a single conscious being at one time, a human
scientific information gathering and utilizing system, an entire planet,
or an entire region of spacetime), it is a `first-person' observation, a
distinct alternative to any other first-person observation of a different
data set.  Since as first-person observations, the different data sets
are mutually exclusive (the first-person at one time observes only one
data set), their probabilities should add up to one, the normalization
principle expressed by Eq.\ (\ref{norm}).

Hartle and Srednicki implicitly discard the first-person nature of the
observation.  Instead of using the full first-person knowledge, {\it ``We
observe the data set $D$,''\/} they consider only the reduced
third-person knowledge and say, {\it ``All we know is that there exists
at least one such region containing our data.''\/}  Therefore, instead of
calculating the likelihood of our data $D$ as the normalized likelihood
of this data set out of all other possible first-person data sets, they
effectively calculate the (different) probability that this data set
exists in at least one region.  That is, they consider not all possible
data sets $D_j$ in $S$ as the alternatives, but simply the binary
alternatives that our particular data set $D$ occurs somewhere and that
$D$ does not occur somewhere \cite{Sredpriv}.

These two procedures, theirs and mine, are equivalent in usual laboratory
experiments in which only one data set actually occurs (in a single
branch of the Everett many-worlds wavefunction).  Then if any data set
$D_j$ occurs that is different from $D$, $D$ does not also occur, so the
probability that $D$ does not occur is the sum of the probabilities for
all $D_j$ different from $D$.  But in a large enough universe, many
different data sets can all actually occur.  Then the probability that
$D$ does not occur can be much smaller than the sum of the probabilities
for each of the other data sets to occur.

The third-person {\it existence\/} of two different data sets, $D_j$ and
$D_k$ with $j\neq k$, is not mutually exclusive or inconsistent, but the
first-person {\it observation} of two different data sets {\it is\/}
mutually exclusive.  Therefore, the third-person existence probabilities
for the data sets can be different from the first-person observational
probabilities for these same data sets.

Ordinary quantum theory with a complete orthogonal set of projection
operators, or the consistent histories approach with a decoherent set of
class operators \cite{Griffiths,Omnes,GH90,GH93}, is well-suited
for calculating the third-person existence probabilities.  But if one
wants the first-person observational probabilities, one needs something
more.

For example, consider a toy model for $S$ that consists of only two
possible data sets, $D_1$ and $D_2$.  Suppose that the quantum state is
such that with unit probability, there exists 1 region with an observer
observing the data set $D_1$, and 999 regions that each have an observer
observing the data set $D_2$.  Ordinary quantum theory would give the
third-person existence probability of both $D_1$ and $D_2$ as unity. 
Since these two existence possibilities are not mutually exclusive, their
existence probabilities do not add up to 1 but rather to 2.  On the other
hand, it would be quite reasonable to assume that the first-person
observational probabilities are the same for all 1\,000 regions, so that
the normalized probability of $D_1$ is 0.001 and of $D_2$ is 0.999.  That
is, there are 999 times as many regions with $D_2$ as there are with
$D_1$ (and we assume that there is nothing else of importance, other than
these different data sets, distinguishing the observers in the two
regions, so all of them can be considered to have equal weight), so the
probability of observing $D_2$ is 999 times the probability of observing
$D_1$.

If the data sets are conscious perceptions, then one way of getting
normalized probabilities for each of them is by the framework of Sensible
Quantum Mechanics \cite{SQM1,SQM2,Page-in-Carr,SQM0,SQM3,SQM4} or
Mindless Sensationalism \cite{MS}, which in the discrete normalizable
case assigns a probability to each conscious perception that is the
expectation value of a corresponding positive `awareness operator.' 
There is no requirement that these positive operators be orthogonal to
each other or even be proportional to projection operators (though they
might be approximately proportional to the integral over all of spacetime
of projection operators in local regions).  In the example above,
assuming that the operator corresponding to $D_1$ (for the first region)
and to $D_2$ (for the remaining 999 regions) receives the same
contribution to the expectation value from each region, then since there
are 999 times as many regions giving $D_2$, the corresponding awareness
operator would have 999 times the probability as that for $D_1$, leading
to the same probabilities as in the previous paragraph.

Hartle and Srednicki \cite{HS} object that calculations like this one
``make the {\it selection fallacy\/} that we are randomly chosen from a
class of objects by some physical process, despite the absence of any
evidence for such a process,'' further stating in a particular example,
{\it ``In fact, there has been no selection at all.''\/}  As mentioned in
the Introduction, I do agree with them that ``We have data that we exist
in the universe, but we have no evidence that we have been selected by
some random process,'' but I disagree with their conclusion that ``We
should not calculate as though we were.''  If the universe does have many
observers, there is indeed no physical selection within them of which
exist and which do not, since they all exist in the third-person sense. 
However, to interpret one's first-person experience, it is perfectly
legitimate to calculate {\it as if\/} it were randomly selected from the
set of all observations.

Bostrom has cogently argued in \cite{Bostrom}, p.\ 162, for the {\it
Strong Self-Sampling Assumption} (SSSA):  ``One should reason as if one's
present observer-moment were a random sample from the set of all
observer-moments in its reference class.'' This is similar to how I might
today \cite{Page06d} state my {\it Conditional Aesthemic Principle} (CAP)
\cite{SQM1}:   ``Unless one has compelling contrary evidence, one should
reason as if one's conscious perception were a random sample from the set
of all conscious perceptions.'' I would argue \cite{Page06d} that the
reference class of all observer-moments (which I would call conscious
perceptions, each being all that one is consciously aware of at once)
should be the universal class of all observer-moments.

Comments analogous to that about the ``selection fallacy'' may be made
about the different branches of the wavefunction in the many-worlds
interpretation of quantum theory, in which the wavefunction never
collapses.  All of the branches with nonzero amplitude may be considered
actually to occur, with no real physical selection between them, but for
an observer predicting what he may observe in the future, it may be
legitimate to make what might be called the {\it Copenhagen fallacy\/}
and calculate as if there were probabilities for the various branches of
the wavefunction to be selected, say by a postulated collapse of the
wavefunction.  Just as in this many-worlds case where it may be
legitimate to reason as if there are probabilities for the selection of a
particular branch of the wavefunction (even if in fact there is no such
selection), so in the many-observations case it may be legitimate to
reason as if there are probabilities for the selection of a particular
observation.

Let me make the parallel between the Copenhagen fallacy and the selection
fallacy more explicit:

Collapse of the wavefunction is false (the ``Copenhagen fallacy'').   But
we can calculate likelihoods as if it happens and use them in a Bayesian
analysis to get posterior probabilities for theories.

Selection of observers is false (the ``selection fallacy''). But we can
calculate likelihoods as if it happens and use them in a Bayesian
analysis to get posterior probabilities for theories.

\section{Consequences}

When the full first-person information about an observation or observed
data set is taken into account (``We observe $D$''), and not just the
third-person account (``$D$ exists''), then we can consider all the
different data sets to be mutually exclusive and hence have normalized
likelihoods.  It is then natural to have likelihoods that vary
monotonically with the typicality assigned to the observation, so that
less typical observations have lower likelihood.  More simply, atypical
observations are unlikely.

The requirement that the likelihoods be normalized means that it does
matter what other observations are possible in a theory, besides what we
may actually observe.  A theory that predicts a huge number of other
possible observations of significant relative likelihood, say by
Boltzmann brains
\cite{Page06a,BF06,Page06b,Linde06,Page06c,Vil06,Page06d,Banks07}, would
tend to give a lower likelihood for our observation than a theory that
does not.  This contradicts the second sentence of Hartle and Srednicki's
conclusion (ii):  ``What other observers might see, how many of them
there are, and what properties they do or do not share with us are
irrelevant for this process.''

Requiring the likelihoods of observations to be normalized first-person
probabilities, instead of the third-person existence probabilities, also
releases theories from the enormous limitations of the second sentence of
Hartle and Srednicki's conclusion (iv):  ``Cosmological models that
predict that at least one instance of our data exists (with probability
one) somewhere in spacetime are indistinguishable no matter how many
other exact copies of these data exist.''  If one were forced to abide by
that limitation, then a huge variety of cosmological models with a
sufficiently large universe (spatially noncompact cosmologies, and also
spatially compact cosmologies with enough inflation) would give nearly
unit probability for our data set and hence the same likelihoods.  Thus
observations would count for nothing in distinguishing between these
theories, and much of cosmology would cease to be an observational
science.

Carter \cite{Carterpriv} has noted that the assumptions of Hartle and
Srednicki, considering only our data and not what other observers might
see, is an example of what, in comparison with the anthropic principle of
assuming that we are typical until it is shown otherwise
\cite{Carterpriv}, he has labeled \cite{Carter06} ``the more sterile and
restrictive autocentric principle.''  Since Hartle and Srednicki's
arguments imply that one could not distinguish observationally any
cosmological theory that gives unit (or even any other equal) likelihood
to our observed data, it certainly seems better to choose other
principles that lead to varying likelihoods and hence the possibilities
of testing cosmological theories observationally.

If Hartle and Srednicki's assumptions were adopted, then theories with a
sufficiently vast and varied multiverse to predict the existence of our
data with near certainty would all have the same weight in a Bayesian
analysis, greater than that of any theory that predicted the existence of
our data with significantly less than unit likelihood.  This would seem
to give an unfair advantage to multiverse theories.  It would also make
them subject to the criticism that they explain everything (since a huge
variety of data sets would then have nearly unit existence likelihood)
and thereby explain nothing.  This seems far too cheap a solution to the
goal of science to explain our observations.

A suitable multiverse theory might turn out to be the best explanation of
our observations, but it should have to earn that status by its high
prior probability (from such considerations as being ``simple, beautiful,
precisely formulable mathematically, economical in their assumptions,
comprehensive, unifying, explanatory, accessible to existing intuition,
etc.\ etc.,'' as Hartle and Srednicki nicely put it) and by not too low a
nontrivial value it gives for the likelihood of our observed data set. 
Replacing the normalized first-person observational probabilities with
the third-person existence probabilities is a cheat, like putting the
theory on steroids.

In a Bayesian analysis to try to find a theory with the maximum {\it a
posteriori\/} probability, it seems unlikely that one can avoid the
tension between trying to make the {\it a priori\/} probability high and
trying to make the likelihood of our observations also high.  For me, the
simplest theory, with the highest {\it a priori\/} probability, would be
the theory that nothing exists.  However, the likelihood of our
observations would then be zero, so this theory is ruled out
observationally.  (It would also run into the problem of not obeying the
normalization principle, since it would give no nonzero observational
likelihoods at all to normalize.)  The theory that everything existed
would seem to me to be the next simplest theory and hence have the next
highest {\it a priori\/} probability.  If one then used existence
probabilities (conditional upon the theory) as likelihoods, as Hartle and
Srednicki seem to advocate, then this simple theory would give unit
likelihoods for all possible observations and hence presumably the
highest {\it a posteriori\/} probability.

Should we then quit physics and say that we have the best possible theory
of everything, namely the simple theory that everything possible exists? 
I would say that this is far too cheap an answer.

If instead we include the normalization principle as I am advocating,
then one would have to normalize the likelihoods of all the
observations.  If, in the theory that everything exists, one made the
simple assignment that all of the infinite number of possible
observations have equal likelihood, then their normalized value would be
zero, and the resulting {\it a posteriori\/} probability for this theory
would be zero, as indeed I would say it is.  One could of course try to
go to an improved theory in which although all possible observations
exist, they have varying likelihoods that are normalized.  Then we are
back to the problem of assigning nontrivial likelihoods, which
complicates the theory and reduces its {\it a priori\/} probability. 
Thus we have the challenge of finding the best theory that neither is so
complicated that it makes its {\it a priori\/} probability too small, nor
has the normalized likelihoods spread so thinly that it makes the
likelihood of our observation too small (e.g., by making us highly
atypical).

One might try to go to the other extreme, maximizing the typicality of
our observed data set by formulating the theory to predict that data set
and only that data set, thereby giving it unit likelihood.  However, it
would be very surprising if any theory existed that predicted our
observations uniquely and was fairly simple.  Since such a theory is
likely to be quite complicated, it would naturally be assigned a low {\it
a priori\/} probability.  Most scientists would presumably believe that
even if one has to reduce the likelihood and typicality of our observed
data from unity, one can gain far more in the {\it a priori\/}
probability for a simpler theory.  In other words, it seems improbable
that only our observed data exists or has nonzero probability, and much
more probable that the correct theory predicts non-unit first-person
observational probability for our data.

If we postulate that the first-person observational probabilities that a
theory predicts are not true probabilities for an actual selection of our
data from all possible data sets, but rather measures for the actual
existence of the various data sets, then giving up on finding a simple
theory predicting unit likelihood for our data set is equivalent to
saying that other data sets actually do exist.  In this way we are led to
a many-observations theory.  The many might be provided by a sufficiently
large universe, by the many-worlds of the Everett version of quantum
theory, and/or by a string landscape.  It seems that the trade-off
between {\it a priori\/} probabilities and likelihoods suggests that many
different observed data sets exist, but not all possible observed data
sets exist equally (i.e., with equal measures or equal likelihoods of
being observed).

\section{Example}

Let us take some set $S$ of all possible data sets under consideration
and consider theories to explain one of them.  Suppose that the set of
all possible data sets is countable (though logically it need not be, if
for example they form a continuum).  Imagine that there is a procedure
for ordering them by their complexity, so that $D_1$ represents the
simplest possible data set, and so on.  Then $D_j$ is the $j$th simplest
data set.  Let us assume that our observed data set has $j \gg 1$, so
that what we observe is by itself not extremely simple.

The theory, say $T_1$, that gives the maximum likelihood for this data
set would be the one that predicts that it alone occurs uniquely, so that
it has unit likelihood, $P(D_j|T_1) = 1$.  Assuming the background
knowledge of the $S$ of all the possible data sets and their ordering by
complexity, theory $T_1$ could be specified simply by giving the integer
$j$.  For most integers $j$ of similar value, this information would not
be compressible, so one could say that the number of bits of information
in this single-observation theory is roughly $\log_2{j}$.

Next, consider an alternative multi-observation theory in which the
likelihoods for the various possible data sets come from some specific
normalized probability distribution.  For simplicity and concreteness,
consider theory $T_N$ which gives the geometric distribution with mean
$N>1$, so $P(D_j|T_N) = (N-1)^{j-1}/N^j$.  If we wanted to choose $N$ to
maximize the likelihood $P(D_j|T_2)$, we would need to choose $N=j$. 
However, this theory, $T_j$, would have more information than $T_1$ (with
the extra amount, above that specifying $N=j$, saying that $T_j$ gives a
geometric distribution).  Hence this $T_j$ theory would presumably be
assigned a lower {\it a priori\/} probability than $T_1$.  Furthermore,
it would also give a lower likelihood, $P(D_j|T_j) = (j-1)^{j-1}/j^j
\approx e^{-j}/j \ll 1 = P(D_j|T_1)$, where the approximation and strong
inequality occurs for $j \gg 1$, as I am assuming.  Therefore, $T_j$
would give a much lower {\it a posteriori\/} probability than $T_1$,
showing that multi-observation theories need not be better than
single-observation theories.

On the other hand, $T_N$ can be chosen to be much simpler than $T_j$
(assuming the generic case in which $j$ is an incompressible large
integer, with roughly $\log_2{j}$ bits of incompressible information) by
choosing $N$ to be much simpler than $j$.  However, to keep the
likelihood $P(D_j|T_j)$ from becoming too much smaller than the maximum
value $P(D_j|T_j) \approx e^{-1}/j$ for fixed $j$ and $N$ allowed to
vary, $N$ should be chosen to be roughly the same value as $j$, say
within a factor of 2 of $j$.  For example, if $N \approx j/2$, then
$P(D_j|T_N) \approx (2/j)e^{-2} \approx (2/e)P(D_j|T_j) \approx 0.736
P(D_j|T_j) \approx 0.271/j$, whereas if $N \approx 2j$, then $P(D_j|T_N)
\approx (0.5/j)e^{-0.5} \approx (\sqrt{e}/2)P(D_j|T_j) \approx 0.824
P(D_j|T_j) \approx 0.303/j$.  Thus for $N$ within a factor of 2 of $j$,
we always get $P(D_j|T_N) > 1/(4j)$.

Now we can simply choose $N$ to be the nearest power of 2 less than or
equal to $j$, $N = 2^{[\log_2{j}]}$, with the square bracket denoting the
integer part of the logarithm to base 2.  Then in binary, $N$ has a 1
followed by $[\log_2{j}]$ 0's, the same as $j$ with all binary digits
after the first truncated to 0.  Thus whereas specifying $j$ requires all
$[\log_2{j}]$ binary digits after the leading 1 to be specified, with
$[\log_2{j}]$ bits of information, $N$ just requires a specification of
how many 0's it has after the leading 1, which is just 
$[\log_2{[\log_2{j}]}]$ bits of information.  For very large generic
(incompressible) $j$, $N$ thus has much less information than that in $j$
itself.

Therefore, if the gain in the {\it a priori\/} probability of $T_N$ from
its relative simplicity of $N$, over that of the more complex $T_1$,
overcomes the decrease in the likelihood of the observed data set $D_j$
from unity for $T_1$ to near $1/(4j)$ for $T_N$, then in Bayes' theorem,
Eq.\ (\ref{bayesthm}), the {\it a posteriori\/} probability $P(T_N|D_j)$
for the multi-observation theory $T_N$ with $N = 2^{[\log_2{j}]}$ will
exceed $P(T_1|D_j)$ for the single-observation theory $T_1$.  In
particular, this would be the case if the prior probabilities obey the
inequality $P(T_N) > 4j P(T_1)$.

Suppose that one re-orders the $T_1$, $T_N$, and other possible theories
(assumed to be countable) into increasing order of complexity and lets
$I$ be the integer that gives this new order, from 1 for the simplest
theory, on up through successively larger integers for more complex
theories.  Then $T_1$ and $T_N$ for integers $N>1$ will all have places
in this order, so that one will get the function $I(i)$ where $i=1$ for
$T_1$ and $i=N$ for $T_N$.  (Of course, the resulting infinite countable
set of values $I(i)$ will not exhaust the positive integers, since there
will also be another countably infinite set of other theories whose $I$'s
will partially intertwine the $I(i)$'s.)  $I$ will be roughly 2 to the
power of the number of bits needed to specify the theory, so $I(1) \sim
j$ and $I(N) \sim \log_2{j}$ for $N = 2^{[\log_2{j}]}$ and $j\gg 1$.

Now let us suppose that we take the {\it a priori\/} probabilities
$P(T_i) \equiv p(I)$ to be a monotonically decreasing function of the
order of complexity $I$, so that simpler theories are assigned higher
prior probabilities and more complex theories are assigned lower prior
probabilities.  If the prior probabilities as a function of $I$, $p(I)$,
fall off too slowly with $I$, then one will get that the posterior
probability of the single-observation theory $T_1$ is greater than that
of the multi-observation theory $T_N$ for any $N$, such as the simple
choice $N = 2^{[\log_2{j}]}$.  However, if $p(I)$ falls off sufficiently
rapidly with $I$, then instead the simpler multi-observation theory will
be favored with the higher {\it a posteriori\/} probability.

In the example above, it appears that it is sufficient for $p(I)$ to fall
off at least as rapidly as $I^{-s}$ for any $s>1$, or even as
$I^{-1}(\ln{I})^{-s}$ for any $s>1$.  For example, consider $p(I) =
(6/\pi^2)I^{-2}$, which would give a normalized prior probability
distribution for a countably infinite set of theories ordered by
complexity, with $I=1$ for the simplest, etc.  This set of priors would
then give the ratio of the posterior probability of the multi-observation
theory $T_N$ to that of the single-observation theory $T_1$ as
$P(T_N|D_j)/P(T_1|D_j) \sim j/(4\log_2{j}) \gg 1$.  Thus with this choice
of priors, the greater simplicity of the multi-observation theory over
the single-observation theory would more than compensate for the reduced
likelihood it gives to the observed data set, so in the end the
multi-observation theory is favored.

Another simple set of prior probabilities that I have advocated
\cite{SQM1,SQM2,Page-in-Carr} is $p(I) = 2^{-I}$.  Since this very
strongly favors simpler theories (each of which is twice as probable {\it
a priori\/} as the next simplest), in the example above it gives a much
higher posterior probability to the multi-observation theory: 
$P(T_N|D_j)/P(T_1|D_j) \sim 2^j/(4j^2)$.

The situation is somewhat similar to theories of solipsism versus
theories in which other people are real.  Solipsism would give a higher
likelihood for one's observations, but it is not nearly so simple as
theories that other people are real.  Therefore, when one chooses prior
probabilities falling sufficiently rapidly with complexity (as humans
apparently do implicitly without even consciously thinking about it), in
the end one favors theories in which other people are real.

The example above shows that typicality itself, in the form of increased
likelihoods, often is not sufficient to overcome the higher prior
probabilities one might like to assign to simpler theories that may
predict larger numbers of possible observed data sets and correspondingly
lower likelihoods for each.  However, this does not work if the simpler
theory predicts too large a range of observed data sets and hence makes
the normalized likelihood of each one too small.  In particular, theories
that predict that all possible observations, out of an infinite set,
occur with equal likelihood give zero likelihood for any particular data
set and hence have zero posterior probabilities (unless absolutely all of
the prior probability is concentrated upon such theories).

\section{Conclusions}

We have seen that when one imposes the {\it normalization principle\/}
and restricts to theories that each give likelihoods summing to unity for
all possible data sets, typicality is automatically favored in the
likelihoods.  Since this preference comes directly from the normalized
likelihoods, it is not and need not be introduced ``through a suitable
choice of priors'' as Hartle and Srednicki \cite{HS} suggest.  Instead,
the prior probabilities for theories may be chosen to ``favor theories
that are simple, beautiful, precisely formulable mathematically,
economical in their assumptions, comprehensive, unifying, explanatory,
accessible to existing intuition, etc.\ etc.,'' as Hartle and Srednicki
propose.

The only sense in which I could be said to favor putting typicality into
the priors would be the interpretation that imposing the normalization
principle effectively assigns zero prior probabilities to theories in
which the likelihoods of all possible observations do not sum to unity,
as I would indeed do if that interpretation were forced upon me.  But not
imposing this requirement does not seem to me to make sense (and also
leads to many sterile cosmological theories that cannot be tested against
observations).  It seems rather analogous to not imposing the requirement
of mathematical consistency.  Therefore, I would argue that the
normalization principle is a fundamental principle of probability for
multi-observation theories that need not be listed among the optional
properties Hartle and Srednicki have nicely enumerated for the
theoretical prejudice of choosing the priors.

Typicality by itself does not guarantee that the theory with the highest
posterior probability will make us typical.  However, typicality is
favored in the likelihoods.  One need not impose it separately, but in
discussions in which one does not explicitly invoke the full Bayesian
framework, assuming typicality may be a legitimate shortcut for selecting
between different theories for our observations.  We are unlikely to be
highly atypical.

\begin{acknowledgments}

I am most grateful to James Hartle and Mark Srednicki for a multitude of
email communications.  The hospitality of the George P.\ and Cynthia W.\
Mitchell Institute for Fundamental Physics of the Physics Department at
Texas A\&M University, and of the Beyond Center for Fundamental Concepts
in Science of Arizona State University, were appreciated for offering me
opportunities to speak in person on this subject with Hartle (and with
other persons such as Gary Gibbons and Seth Lloyd).  The hospitality of
Edgar Gunzig and the Cosmology and General Relativity symposium of the
Peyresq Foyer d'Humanisme in Peyresq, France, enabled me to talk to
Brandon Carter and others there about these ideas.  This research was
supported in part by the Natural Sciences and Engineering Research
Council of Canada.

\end{acknowledgments}

\end{document}